\documentclass[conference]{IEEEtran}
\IEEEoverridecommandlockouts

\usepackage{cite}
\usepackage{amsmath,amssymb,amsfonts}
\usepackage{graphicx}
\usepackage{textcomp}
\usepackage{xcolor}
\usepackage{adjustbox}
\usepackage{caption}
\usepackage{subcaption}
\usepackage{array}
\usepackage{xspace}

\usepackage{multirow}

\usepackage{array} 
\newcolumntype{L}[1]{>{\raggedright\arraybackslash}p{#1}}

\usepackage{algorithm}
\usepackage[noend]{algpseudocode}
\usepackage{tikz}
\usepackage[hidelinks]{hyperref}

\algrenewcommand\algorithmiccomment[1]{\hfill{\(\triangleright\) #1}}
\algnewcommand\algorithmicinput{\textbf{Input:}}
\algnewcommand\algorithmicoutput{\textbf{Output:}}

\def\BibTeX{{\rm B\kern-.05em{\sc i\kern-.025em b}\kern-.08em
    T\kern-.1667em\lower.7ex\hbox{E}\kern-.125emX}}

\newcommand{\framework}{DPUConfig\xspace}

\newcommand\copyrighttext{%
	\footnotesize This is the preprint accepted for publication in the Design, Automation and Test in Europe Conference (DATE 2026), Verona, Italy, 20–22 April 2026. This version is released under a CC-BY license in accordance with the Horizon Europe programme requirements.}
\newcommand\copyrightnotice{%
	\begin{tikzpicture}[remember picture,overlay]
		\node[anchor=north,yshift=-10pt] at (current page.north) {\fbox{\parbox{\dimexpr\textwidth-\fboxsep-\fboxrule\relax}{\copyrighttext}}};
	\end{tikzpicture}%
}
    
\begin{document}

\title{\framework: Optimizing ML Inference in FPGAs Using Reinforcement Learning}

\author{

\thanks{The research is co-funded by the European Union’s Horizon Europe Programme under the MLSysOps Project (Grant Agreement No. 101092912). It is also conducted in the operating framework of the University of Thessaly Innovation, Technology Transfer Unit and Entrepreneurship Center One Planet Thessaly, under the “University of Thessaly Grants for Scientific Publication Support” action and is co-funded by the Special Account of Research Grants of the University of Thessaly.}

\IEEEauthorblockN{
Alexandros Patras,
Spyros Lalis,
Christos D. Antonopoulos,
and Nikolaos Bellas
}
\IEEEauthorblockA{
Department of Electrical and Computer Engineering, University of Thessaly, Volos, Greece\\
\{patras, lalis, cda, nbellas\}@uth.gr}
}

\maketitle

\copyrightnotice

\begin{abstract} Heterogeneous embedded systems, with diverse computing elements and accelerators such as FPGAs, offer a promising platform for fast and flexible ML inference, which is crucial for services such as autonomous driving and augmented reality, where delays can be costly. However, efficiently allocating computational resources for deep learning applications in FPGA-based systems is a challenging task. 
A Deep Learning Processor Unit (DPU) is 
a parameterizable FPGA-based accelerator module optimized for ML inference. It supports a wide range of ML models and can be instantiated multiple times within a single FPGA to enable concurrent execution.
This paper introduces \framework, a novel runtime management framework, based on a custom Reinforcement Learning (RL) agent, that dynamically selects optimal DPU configurations by leveraging real-time telemetry data monitoring, system utilization, power consumption, and application performance to inform its configuration selection decisions. The experimental evaluation demonstrates that the RL agent achieves energy efficiency 95\% (on average) of the optimal attainable energy efficiency for several CNN models on the Xilinx Zynq UltraScale+ MPSoC ZCU102.
\end{abstract}

\begin{IEEEkeywords}
FPGA, adaptive, configuration, reinforcement learning, power efficiency. 
\end{IEEEkeywords}

\section{Introduction}
\label{intro}

Recent advances in machine learning have driven a surge in the deployment of deep learning inference tasks on heterogeneous computing platforms. While traditional systems have leveraged multi-core CPUs and GPUs, MPSoCs with field-programmable gate arrays (FPGAs) have also emerged as a highly efficient alternative due to their flexibility and energy efficiency. During the past decade, many FPGA-based neural network accelerator designs have been introduced~\cite{mittal2020survey, yan2024surveyfpgabasedacceleratorml}. In particular, Deep Learning Processing Units (DPUs)~\cite{xilinx_dpuczdx8g} offer a promising platform for accelerating ML workloads.

A key challenge in leveraging MPSoCs with FPGAs for deep learning inference lies in the inherent variability of workloads and the diverse configuration options available. Unlike conventional CPU scheduling —-where the focus is primarily on allocating processing cores to balance latency and power consumption—- FPGA platforms offer additional degrees of freedom. These include selecting among various DPU configurations and choosing different hardware parameters tailored to the specific requirements of each inference task. The configuration options can greatly affect not only the inference performance but also the energy efficiency. In MPSoCs, the CPU also plays a crucial role in managing FPGA kernel execution, directly impacting system responsiveness.

To address these challenges, this work proposes an adaptive scheduling framework based on Reinforcement Learning (RL), 
which integrates real-time telemetry data, such as CPU and memory system utilization, power dissipation, and application performance, into its decision-making process. By harnessing a custom RL agent, our system dynamically determines the optimal task-to-hardware mapping, choosing the most suitable DPU configuration for each incoming inference task. This enables the runtime system to balance the trade-offs between computation latency and power dissipation while accommodating the unique constraints imposed by FPGA architectures.

This paper makes the following key contributions:
\begin{itemize}
    \item It introduces \framework, a custom RL-based run-time system that, under stochastic variability and user constraints, consistently chooses DPU configuration for ML inference, which is very close to the optimal one. Although prior work has applied RL-based resource allocation or configuration to optimize specific metrics, this is the first work to apply those methods to reconfigurable DPUs. 
    \item To demonstrate the feasibility and practicality of \framework, we implement and evaluate it with a variety of ML inference use cases on the ZCU102 device and show that the benefits of using this agent for DPU configuration decisions outweigh potential reconfiguration overheads.
\end{itemize}

\section{Deep Learning Processor Units (DPUs)}
\label{dpu}
Deep Learning Processor Units (DPU)~\cite{xilinx_dpuczdx8g}, released as part of the Xilinx Vitis-AI toolchain~\cite{amd2023vitisai}, have become a popular solution to deploy pre-trained ML models on FPGA devices. 
DPUs are programmable, featuring a CISC-style instruction set capable of supporting inference for a wide range of CNN models.
DPUs are used in an increasingly large number of applications such as autonomous driving~\cite{wu2019autonomous}, object detection~\cite{Li2023EdgeRealTimeDPU, Amin2024FPGARealTime}, thermal imaging~\cite{hussein2022thermal}, and even semantic segmentation for space applications~\cite{Perryman2025DependableDPU}.

Multiple instances of DPUs can be used to run independent ML inferences concurrently (Table~\ref{tab:dpu_config}). For example, due to resource constraints on the ZCU102 MPSoC, only up to three large B4096 DPUs can be instantiated to execute three independent ML models. The B4096 configuration has Pixel Parallelism (PP) = 8 and Input and Output Channel Parallelism (ICP) = (OCP) = 16. Hence, the peak performance is 2048 MAC operations per cycle~\cite{xilinx_dpuczdx8g}. Each MAC operation counts for two regular operations. 

For preparing the necessary artifacts, the Vitis AI framework reads the pre-trained ONNX~\cite{onnx} or PyTorch~\cite{pytorch} representation of the ML model, optionally performs pruning and INT8 quantization, and then compiles the quantized model for a target DPU architecture using the Vitis AI compiler~\cite{amd2023vitisai}. To exploit parallelism within the layers of the ML model, the compiler performs multiple optimizations, such as layer fusion and instruction scheduling. Once execution starts, the DPU fetches its instructions from DDR memory, while on-chip BRAM/DRAM buffers store the inputs, outputs, and intermediate data to reduce latency and minimize external memory bandwidth usage. The DPUs are invoked by the host CPU and execute the CNNs layer by layer.

\begin{table}[h]
    \caption{Name, maximum number of DPU instances, and the selected configurations used in the action space of the RL agent, based on the DPUCZDX8G IP~\cite{xilinx_dpuczdx8g} for the Zynq UltraScale+ MPSoC~\cite{xilinx_zcu102}. For example, we consider 4 instances in B1600.}
    \label{tab:dpu_config}. 
   \centering
   \setlength{\tabcolsep}{4pt}
   \begin{tabular}{|l|c|l|L{3cm}|} \hline
     \textbf{DPU configuration} & \textbf{Max.} & \textbf{Notation} & \textbf{Selected} \\ 
     \textbf{(PP*ICP*OCP)} & \textbf{instances} &  & \textbf{Configurations} \\ \hline
       B512 (4*8*8)    & 8 & B512\_8   & B512\_\{1,4,8\}\\
       B800 (4*10*10)  & 7 & B800\_7   & B800\_\{1,4,7\} \\
       B1024 (8*8*8)   & 6 & B1024\_6  & B1024\_\{1,3,6\}\\
       B1152 (4*12*12) & 6 & B1152\_6  & B1152\_\{1,3,6\} \\
       B1600 (8*10*10) & 4 & B1600\_4  & B1600\_\{1,2,3,4\} \\
       B2304 (8*12*12) & 4 & B2304\_4  & B2304\_\{1,2,3,4\} \\
       B3136 (8*14*14) & 3 & B3136\_3  & B3136\_\{1,2,3\} \\
       B4096 (8*16*16) & 3 & B4096\_3  & B4096\_\{1,2,3\} \\ \hline
   \end{tabular}
\end{table}

\section{Motivation}
\label{motivation}

This section presents results of system characterization for realistic deep neural network (DNN) inference scenarios using DPU acceleration and the Vitis AI framework on the Xilinx Zynq Ultrascale+ FPGA board (ZCU102)~\cite{xilinx_zcu102}. Our analysis explores the design space along three key dimensions, namely latency, accuracy, and energy efficiency (measured as performance per watt, PPW), to identify the most critical features that determine the optimal DPU configuration for efficient ML inference.

\subsection{The optimal DPU configuration depends on the characteristics of the ML model.}

\begin{figure}[!h]
  \centering
  \includegraphics[width=\linewidth]{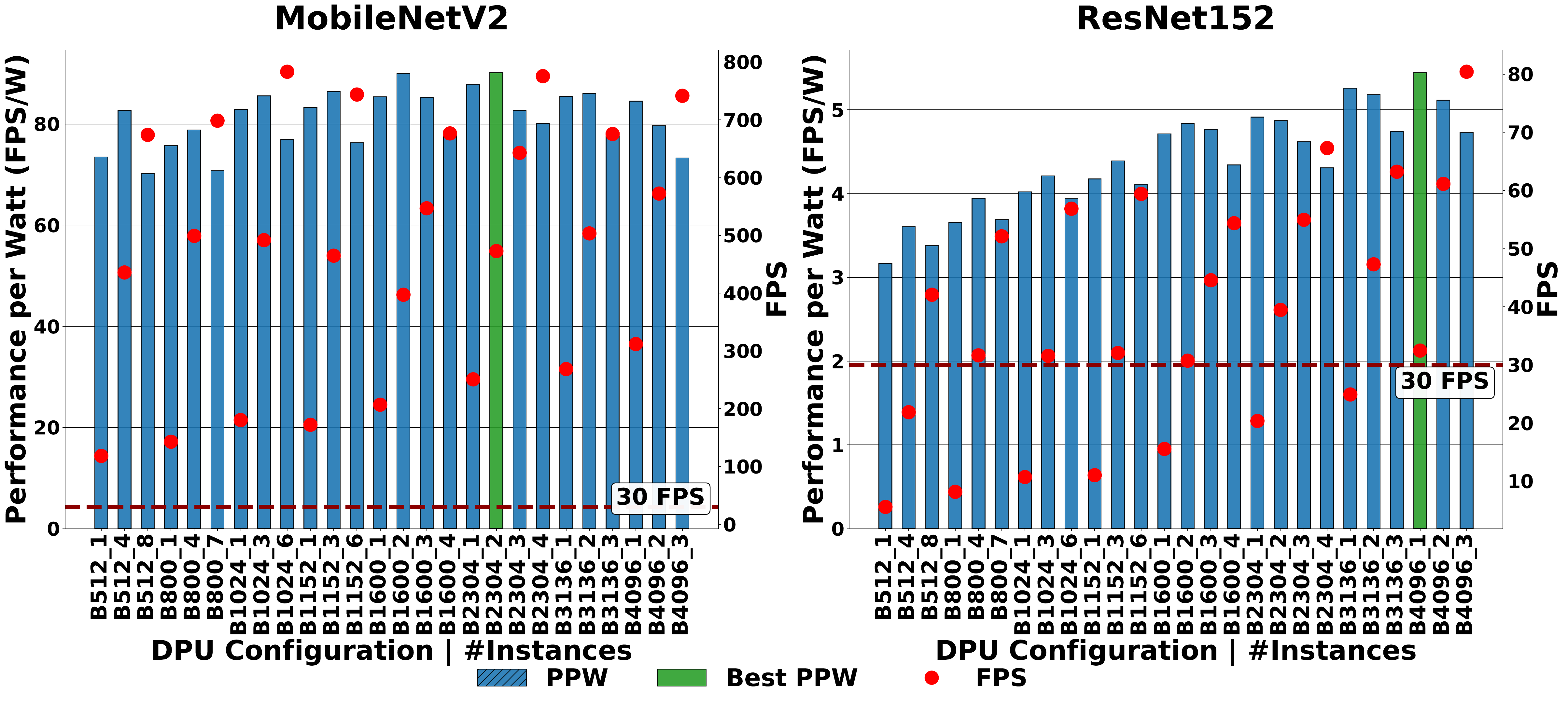}
 \caption{The optimal execution target depends on ML characteristics. The bars (left axis) show energy efficiency in FPS per Watt, and the red points (right axis) indicate performance.}     
 \label{fig:ppw}
\end{figure}

Fig.~\ref{fig:ppw} shows the energy efficiency (PPW) and the performance (in Frames Per Second) of two representative ML inference use cases over different DPU configurations when a single ML model type runs on the DPUs. Considering only DPU configurations that deliver at least 30 fps, we observe that the optimal DPU configuration depends on the characteristics of the model. For example, with ResNet152~\cite{He2016ResNet}, the B4096\_1 configuration offers the highest energy efficiency; meanwhile, for MobileNetV2~\cite{Sandler2018MobileNetV2}, B2304\_2 yields the highest efficiency. 

Even if larger DPU size configurations provide the highest throughput and energy efficiency for many ML models, this is not the case for smaller models with lower arithmetic intensity and lower DPU utilization, such as MobileNetV2 (see Table~\ref{tab:model_performance}). Such models do not utilize all the resources of a larger DPU (DPU utilization is only 17.1\%), but they seem to benefit from more instances of smaller DPUs. The MobileNetV2 performance in B4096\_1 is only 2.6x higher than in B512\_1. In contrast, ResNet152, being more compute-bound, achieves a much larger 5.8x speedup under the same settings.

\subsection{CPU interference from co-executing applications may alter the optimal DPU configuration.}

To study how concurrent external workloads in the multicore CPU affect execution variability, artificial CPU- and memory-intensive workloads are introduced, resulting in three system states (N, C, M): (i) None
for no additional workload (N), (ii) computation-intensive workloads minimally using memory bandwidth (C), and (iii) 
memory-intensive workloads that continuously maintain high memory bandwidth utilization (M).

\begin{figure}[!h]
  \centering
  \includegraphics[width=\linewidth]{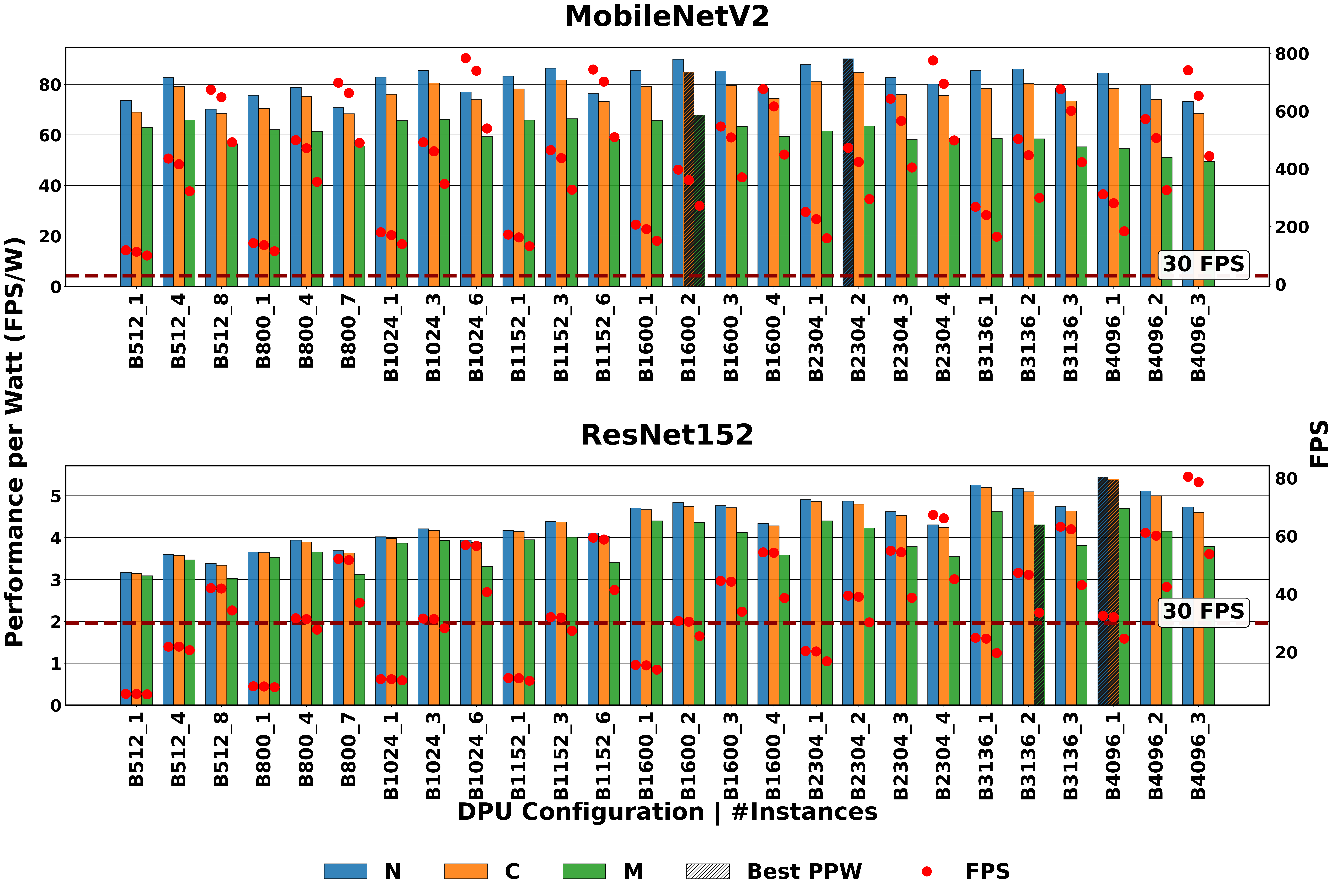}
    \caption{PPW (left axis, bars) and performance in FPS (right axis, points) across different DPU configurations under three system states. The dark bars highlight the configuration achieving the best energy efficiency while maintaining performance above 30 FPS.} 
  \label{fig:workload_figure}
\end{figure}

Fig.~\ref{fig:workload_figure} indicates that, in order to meet the 30 fps requirement, the most energy efficient DPU configuration for MobileNetV2 is B1600\_2 in the C and M states; by contrast, for the N state, the most efficient setup remains B2304\_2. DPU performance is strongly influenced by the memory bandwidth utilization of competing workloads. In such cases, a smaller DPU achieves a better PPW ratio, since larger DPUs are deprived of sufficient bandwidth to reach peak performance and spend more cycles stalled while waiting for data. Furthermore, smaller models exhibit shorter execution times (see Table~\ref{tab:model_performance}), which increases the activity of the CPU thread responsible for coordinating DPU execution. This makes them more susceptible to higher response latencies under heavy CPU load. ResNet152 demonstrates a similar trend, with the best PPW in memory-intensive state M achieved by a smaller DPU configuration (B3136\_2).

\subsection{The optimal DPU configuration varies with inference accuracy requirements.}

\begin{figure}[!h]
  \centering
  \includegraphics[width=\linewidth]{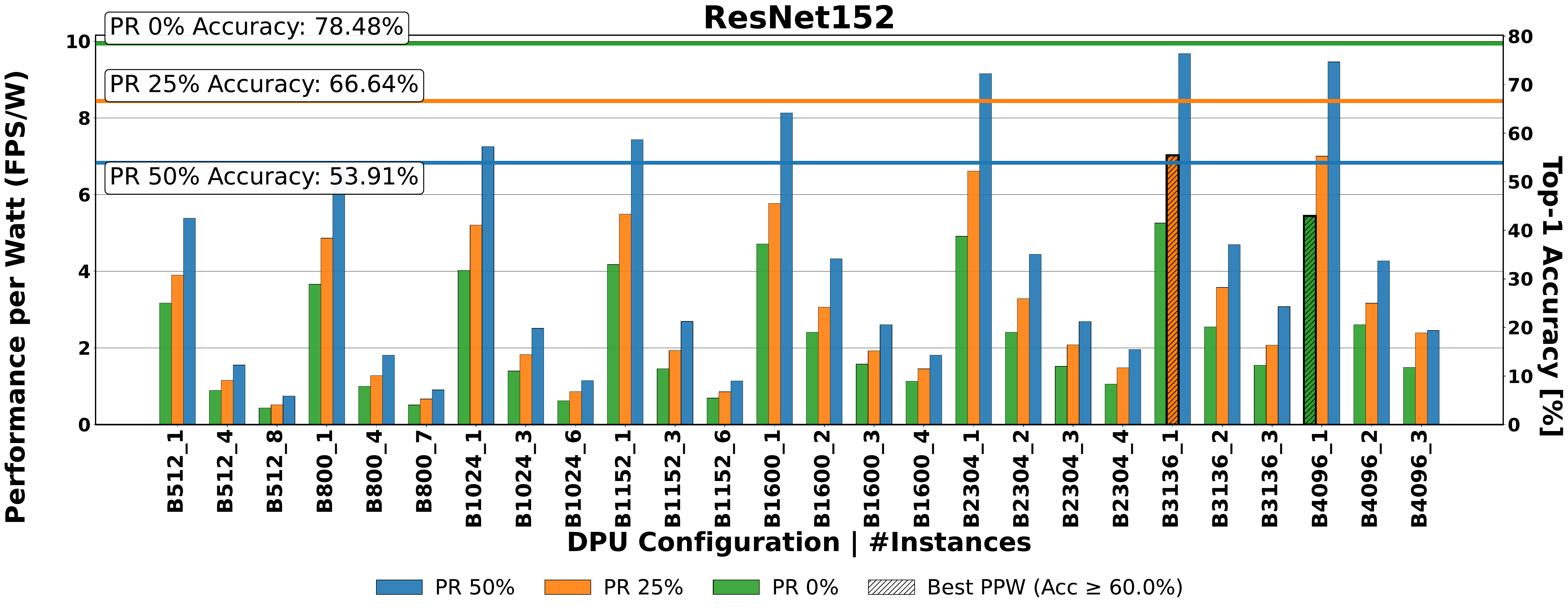}
 \caption{PPW (left axis, bars) and accuracy (right
axis, lines) across different DPU configurations under the N state. For example, the accuracy of ResNet152 when 25\% of its channels are eliminated is 66.64\%. }      
 \label{fig:ppw_accuracy}
\end{figure}

Vitis AI employs channel pruning, which removes entire channels/filters in convolutional layers~\cite{EagleEye2020}. This reduces model size and increases performance, but at the cost of reduced accuracy. Fig.~\ref{fig:ppw_accuracy} presents the energy efficiency for three versions of ResNet152, corresponding to pruning ratios of 0\%, 25\%, and 50\%. For an accuracy threshold of 60\%, ResNet152 can be pruned by 25\% to radically improve energy efficiency compared to the original model using a different DPU configuration (B3136\_1 instead of B4096\_1). Thus, the availability of differently pruned model variants, combined with a specific accuracy target, complicates dynamic selection of DPU configurations to maximize PPW while meeting accuracy requirements.

\section{Design of the \framework framework}
\label{rl}

\begin{figure*}[htb]
  \centering
  \includegraphics[width=\textwidth]{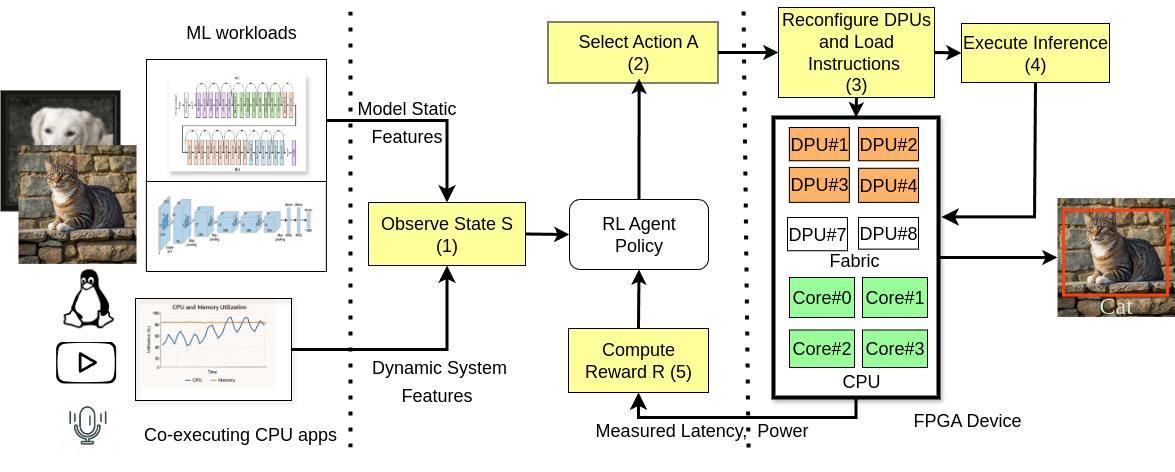}
  \caption{High-level design of the \framework framework.}
  \label{fig:framework}
\end{figure*}

Fig.~\ref{fig:framework} shows the overview of the \framework framework. The core component of \framework is the RL agent, a machine learning approach focused on learning a sequence of decisions to maximize cumulative rewards~\cite{sutton2018reinforcement}. The primary objective of the RL agent is to determine an optimal DPU configuration under specific constraints. 
\framework observes the state S of the system, including the static characteristics of the ML model and the runtime metrics from telemetry monitoring, as shown in Table~\ref{tab:rl_state}. Then, it uses the RL agent to select the next action A that will maximize the energy efficiency of the platform while satisfying the latency constraints set by the user. Action A selects the next DPU configuration (size and number of compute units). \framework then dynamically reconfigures the FPGA with the new DPU configuration, loads the DPU instructions for the ML model, and executes the inference on the new configuration. Based on the measured execution metrics (fps and power), \framework computes the reward function, which evaluates how action A improves the energy efficiency under the latency constraints.  

\subsection{Reinforcement Learning Agent}
\label{subsec:rl}
This section defines the three core components that formulate the optimization space of \framework.

\textbf{State.} Based on the analysis of Section~\ref{motivation}, Table~\ref{tab:rl_state} summarizes the state features that capture both the runtime system status and the model characteristics, which together form the input to the RL agent. The \emph{dynamic system features} include per-core CPU utilization, the usage of memory bandwidth across read and write memory ports, and the power consumption of the FPGA and CPU subsystems. These metrics are collected periodically at runtime.

The \emph{model static features} describe the ML model itself, including the number of MAC operations, the memory load/store requirements, and the total number of parameters (shown in Table~\ref{tab:model_performance}). These static features are known \textit{a priori} and are collected once when a new model is introduced into \framework. Finally, each model specifies its own performance constraints.

\begin{table}[!h]
  \caption{State features.}
  \label{tab:rl_state}
  \centering
  \setlength{\tabcolsep}{4pt}
  \begin{tabular}{|l|p{7cm}|}
    \hline
    \textbf{State} & \textbf{Description} \\ 
    \hline\hline

    \multicolumn{2}{|c|}{\emph{Dynamic System Features}} \\ 
    \hline
    ${CPU_i}$ & Utilization of CPU core $i \in \{0,1,2,3\}$  \\ 
    \hline
    ${MEMR_j}$ & Memory read bandwidth (MB/s) of port $j \in \{0,\dots,4\}$ \\ 
    \hline
    ${MEMW_j}$ & Memory write bandwidth (MB/s) of port $j \in \{0,\dots,4\}$ \\ 
    \hline
    ${P_\mathrm{FPGA}}$ & Power consumption (W) of FPGA fabric \\ 
    \hline
    ${P_\mathrm{ARM}}$ & Power consumption (W) of CPU cores\\ 
    \hline

    \multicolumn{2}{|c|}{\emph{Model Static Features}} \\ 
    \hline
    GMAC & Number of MAC operations (GMACs) \\ 
    \hline
    LDFM & Model load from memory (bytes) \\ 
    \hline
    LDWB & Model load from weight buffer (bytes) \\ 
    \hline
    STFM & Model store to memory (bytes) \\ 
    \hline
    PARAM & Number of trainable model parameters \\ 
    \hline    
    \multicolumn{2}{|c|}{\emph{Constraints}} \\ 
    \hline
    C\_PERF & FPS Performance constraint \\ 
    \hline

  \end{tabular}
\end{table} 

\textbf{Actions.} The available actions correspond to 26 distinct DPU configurations, combining different DPU sizes and numbers of instances, as listed in Table~\ref{tab:dpu_config}. The action space does not cover the entire design space, as certain intermediate configurations (e.g., B512\_2, B800\_6) were omitted. These configurations were excluded based on empirical analysis, which showed that they do not provide meaningful variation in training and are never part of the optimal configuration set.

\textbf{Reward.} Rewards provide the feedback that guides the RL agent toward the desired goal. Each action must contribute to achieving a specific target. In our setting, however, the optimization target is inherently dynamic: the achievable performance-per-watt (PPW) depends on both the current system workload and the ML model characteristics.

This context dependency introduces two challenges. First, there is no global reward target, as an action that maximizes PPW under one workload state may be suboptimal in another. Second, naive training without context awareness risks overfitting to the limited states seen during training, leading to poor generalization for unseen workloads or models. These challenges are related to moving-target problems in RL, studied in contextual bandits~\cite{liContextualBandit}, input-driven environments~\cite{maoVarianceReductionReinforcement2018}, and state-dependent baselines~\cite{tuckerMirageActionDependentBaselines2018}. 

To address this, we adopt a context-aware reward design. Workload-dependent state (CPU and memory port utilization, and model characteristics such as GMACs and data transfers) are included in the state. Rewards are defined relative to context-specific baselines rather than absolute PPW. This ensures that the agent evaluates performance in relation to what is achievable under the current workload and model. By combining contextual baselines with normalized reward shaping~\cite{maoVarianceReductionReinforcement2018, tuckerMirageActionDependentBaselines2018, ngPolicyInvarianceReward1999}, the agent learns policies that generalize across workloads and models while avoiding the moving-target problem.

Algorithm~\ref{algo:reward} formalizes reward calculation using the sampled system metrics. If performance constraints are not met, a negative reward is returned. Otherwise, the reward is based on PPW, normalized against a blended baseline. This baseline is computed from two sources: a local average $b_{\text{local}}$ from the current context bucket (defined by the current workload-dependent state) and a global average $b_{\text{global}}$ across all contexts. A blending factor $\lambda$ interpolates between them to balance local adaptation and global stability. The local average $b_{\text{local}}$ is updated online with new PPW samples, capturing efficiency under similar conditions, while the global average $b_{\text{global}}$ aggregates statistics across all buckets to provide a fallback when data is sparse. 
The factor $\alpha$ scales the reward to avoid extreme values. This formulation ensures bounded rewards, emphasizes relative improvements, and prevents unstable updates during training.

\begin{algorithm}[t]
\caption{Reward calculation based on PPW and constraints.}
\label{algo:reward}
\begin{algorithmic}[1]
\Procedure{CalculateReward}{$S$}
  \State measuredFPS $\gets S.fps$
  \State fpgaPower $\gets S.fpgaPower$
  \State cpuUtil $\gets S.cpu$, memUtil $\gets S.memory$
  \State gmac $\gets S.gmac$, modelData $\gets S.modelData$
  
  \State ppw $\gets \tfrac{measuredFPS}{fpgaPower}$
  
  \If{$ ( measuredFPS \textless FPSConstraint )$}
    \State $r \gets$ -1.0
    \State \Return $r$
  \EndIf
  
  \State $contextKey \gets (cpuUtil, memUtil, gmac, modelData)$
  \State $b_{\text{local}} \gets \textsc{ctxMean}[contextKey]$
  \State $b_{\text{global}} \gets \textsc{globalMeanPPW}$
  \State baseline $\gets (1-\lambda)b_{\text{local}} + \lambda b_\text{global}$
  \State $r \gets \tanh\!\Big(\dfrac{\text{ppw} - \text{baseline}}{\alpha \cdot \max(1,|baseline|)}\Big)$
  \State Update \textsc{ctxMean}, \textsc{globalMeanPPW}
  \State \Return $r$
\EndProcedure
\end{algorithmic}
\end{algorithm}

In conventional RL, agents maximize cumulative rewards, which are often assumed to increase monotonically. However, in our case, the reward is defined as the relative improvement over a context-specific baseline, and thus it naturally fluctuates around zero: Positive values indicate better-than-baseline performance, while negative values represent degradation. Without bounding, these fluctuations can lead to instability during learning when the difference from the baseline is extreme. Prior work has shown that reward clipping or squashing functions help stabilize training by limiting the influence of outliers~\cite{mnihHumanlevelControlDeep2015} and preventing the policy from overemphasizing rare high-magnitude deviations~\cite{fuRewardShapingMitigate2025}. As a result, the policy learns to generalize across varying workloads and model states without being destabilized by rare outliers, consistent with findings in prior work on stabilized RL training~\cite{hesselRainbowCombiningImprovements2018}.

\textbf{Training.} 
Algorithm~\ref{algo:ppo} summarizes the training procedure of our RL agent. 
Instead of running live hardware experiments during training, we rely on a large set of pre-recorded measurements. These were collected from exhaustive runs covering different DPU configurations, ML models, and workload states. At each training step, the system is initialized to an initial state according to the chosen workload mode (C, N, or M) and target model. We adopt single-step episodes: the agent observes the initial state, selects an action, and the outcome is retrieved from the corresponding pre-recorded experiment data. The collected system metrics and model characteristics are then passed to the reward function (Algorithm~\ref{algo:reward}) to compute the training signal. 
Finally, the Proximal Policy Optimization (PPO)~\cite{schulmanPPO2017} backend updates the policy based on the observed reward. This process is repeated across all state-action combinations, and the resulting trained RL agent is returned.  

\begin{algorithm}[t]
\caption{Training the RL Agent with PPO}
\label{algo:ppo}
\begin{algorithmic}[1]
\Procedure{TrainRLAgent}{}
  \For{episode $=1 \dots N$}
    \State Select workload mode (N, M, C) \& ML model
    \State Initialize system to empty state
    \State Observe initial system metrics and model features
    \State $a \gets $ Agent selects action based on current policy
    \State $state \gets $ Fetch telemetry data
    \State $r \gets $ \textsc{calculateReward} $(state)$
    \State Update PPO policy parameters using $(state,a,r)$
  \EndFor
  \State \Return trained RL agent
\EndProcedure
\end{algorithmic}
\end{algorithm}

\section{Experimental Evaluation}
\label{evaluation}

\subsection{Experimental Setup}
\label{eval:setup}

\begin{table*}[!h]
\caption{ML model characteristics. Latency and Data I/O refer to the inference of a single image using the B4096\_1 configuration. The reported accuracy is for \textit{INT8} quantized models without pruning. For YOLOv5s, accuracy refers to Mean Average Precision (mAP). Each model has also two pruned versions (25\% \& 50\%).}
\centering
\setlength{\tabcolsep}{4pt}
\begin{tabular}{|c|c|c|c|c|c|c|c|c|c|}
\hline
\textbf{Type} & \textbf{Model} & \textbf{Latency} & \textbf{Avg. \textit{INT8}} & \textbf{\# Layers} & \textbf{\# GMAC} & \textbf{Data I/O between} & \textbf{Bandwidth} & \textbf{Arithm. Intensity} & \textbf{DPU} \\
 &  & \textbf{(ms)} &  \textbf{Accuracy} &  & \textbf{operations} & \textbf{DRAM--DPU (MB)} & \textbf{(GB/s)} & \textbf{(MACs/Byte)}  &   \textbf{Efficiency} \\
\hline

\multirow{8}{*}{\textbf{Training}} 
& ResNet18      & 4.43  & \(67.90\%\) & 18  & 1.82 & 12.13 & 2.03 & 149.83 & \(71.90\%\) \\
& ResNet50      & 11.72 & \(77.60\%\) & 50  & 4.10 & 38.94 & 2.85 & 105.33 & \(59.00\%\)  \\
& MobileNetV2   & 3.21  & \(68.23\%\) & 53  & 0.30 & 5.74  & 1.49 & 52.49  & \(17.10\%\) \\
& DenseNet121 \cite{densenet} & 17.39 & \(68.70\%\) & 98  & 2.86 & 43.74 & 2.93 & 65.28 & \(26.90\%\) \\
& InceptionV4   & 32.23 & \(77.14\%\) & 150 & 12.3 & 89.00 & 2.54 & 138.23 & \(63.00\%\) \\
& RepVGG A0     & 4.83  & \(72.41\%\) & 45  & 1.52 & 11.84 & 2.00 & 128.26 & \(53.40\%\) \\
& ResNext-50 32x4d \cite{resnext} & 27.42 & \(76.21\%\) & 50 & 11.41 & 95.85 & 3.17 & 119.06 & \(68.90\%\) \\
& YOLOv5s       & 34.70 & \(42.10\%\)  & 60  & 8.26 & 159.80 & 3.27 & 51.69  & \(42.90\%\) \\
\hline

\multirow{3}{*}{\textbf{Test}} 
& RegNetX 400MF \cite{regnet} & 5.71  & \(70.15\%\) & 72  & 1.57 & 24.33 & 3.76 & 64.57 & \(47.40\%\) \\
& InceptionV3   & 15.03 & \(77.03\%\) & 98  & 5.74 & 43.13 & 2.46 & 133.05 & \(63.50\%\) \\
& ResNet152     & 30.81 & \(78.48\%\) & 152 & 11.54 & 76.52 & 2.35 & 150.81 & \(62.00\%\) \\
\hline
\end{tabular}
\label{tab:model_performance}
\end{table*}

We perform our evaluation on the Xilinx Zynq UltraScale+ FPGA board ZCU102~\cite{xilinx_zcu102}, using the Vitis AI v3.5 framework~\cite{amd2023vitisai}, and the DPU v4.1~\cite{xilinx_dpuczdx8g}. 
Our framework is evaluated using ten CNNs for image classification and the YOLOv5 detector (Table~\ref{tab:model_performance}), which have varying compute and memory demands. For each model, two pruned variants with pruning ratios 25\% and 50\% were also added, for a total of 33 models. The ImageNet~\cite{Deng2009ImageNet} dataset was used as input for the classification networks, and COCO~\cite{lin2015microsoftcococommonobjects} for the object detector. We use a PyTorch~\cite{pytorch}-based script and the \texttt{vaitrace}~\cite{amd2023vitisai} tool to extract static model features, and the \texttt{stress-ng}~\cite{king2025stressng} utility to emulate the workload states C and M.

In total, 2574 experiments were executed, covering the space of models, configurations, and workload states (26 DPU configurations $\times$ 11 models $\times$ 3 pruned variants $\times$ 3 workload states). Each experiment was run for a fixed number of input images. An OpenTelemetry collector~\cite{otel} ran on a separate machine, collecting system metrics exported by a Prometheus Node Exporter~\cite{nodexporter} instance on the ZCU102 at 3Hz, alongside application performance metrics. Power measurements were obtained from the integrated sensors of the ZCU102 board.

The CNN models were split into two sets: i) 24 for training and ii) 9 for testing. The split was determined via k-means clustering on GMAC values, grouping the models into three categories: small, medium, and large. One representative model with its two pruned variants from each category was placed in the test set. All three workload modes (N, C, M) were included during training.
Model and workload combinations were presented to the RL agent in round-robin order, ensuring exposure to the full range of state–action pairs. The RL agent was developed using the OpenAI Gymnasium~\cite{brockman2016openai} environment and trained using Ray RLLib~\cite{raylib}.

\subsection{Results}
\label{eval:res}

\begin{figure*}[!h]
  \centering
  \includegraphics[width=\linewidth]{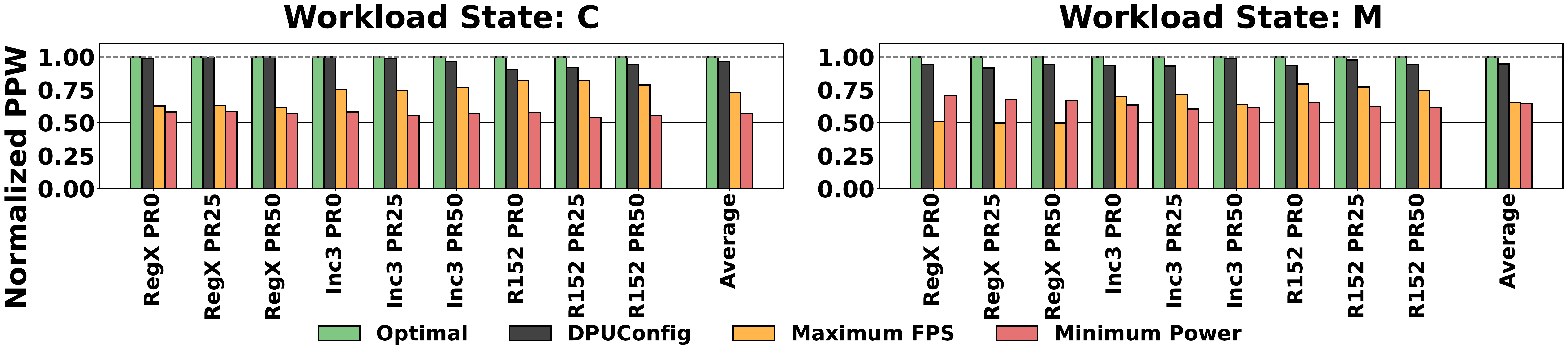}
    \caption{Normalized PPW results of \framework across two workload states (C, M). 
Model Abbr.: RegX = RegNetX, Inc3 = InceptionV3, R152 = ResNet152. PR0, PR25, PR50 denote pruning ratios of 0\%, 25\%, and 50\%, respectively.} 
    \label{fig:workload_fig}
\end{figure*}

Fig.~\ref{fig:workload_fig} presents the normalized PPW achieved by \framework compared against: i) an Optimal configuration, which always runs inference on the best DPU configuration, ii) the configuration with the maximum FPS, and iii) the one with minimum power dissipation.
For the workload state C, \framework achieves on average 97\% of the optimal PPW, with two cases exactly matching the optimal configuration. In workload state M, where memory bandwidth is heavily stressed and DPU performance is further degraded, the average drops slightly to 95\%, with no case reaching the exact optimum. This is a strong indication that the RL agent is tuned to generalize across different models and system states rather than overfitting to specific cases. The maximum-FPS configuration (typically B4096\_1) achieves only 47\% of the optimal PPW in workload state C and 35\% in workload state M, showing that the largest DPU is not necessarily the most energy-efficient. Similarly, the minimum-power configuration (B512\_1) consistently falls far short of optimal efficiency. These results confirm that neither extreme (largest nor smallest DPU) is efficient and that \framework achieves near-optimal energy efficiency in all conditions. In all evaluation experiments, the performance constraint was set to 30~FPS and was satisfied in 89\% of the test cases, with violations occurring only for the demanding ResNet152 model under the M workload state.

\begin{figure}[!h]
  \centering
  \includegraphics[width=\linewidth]{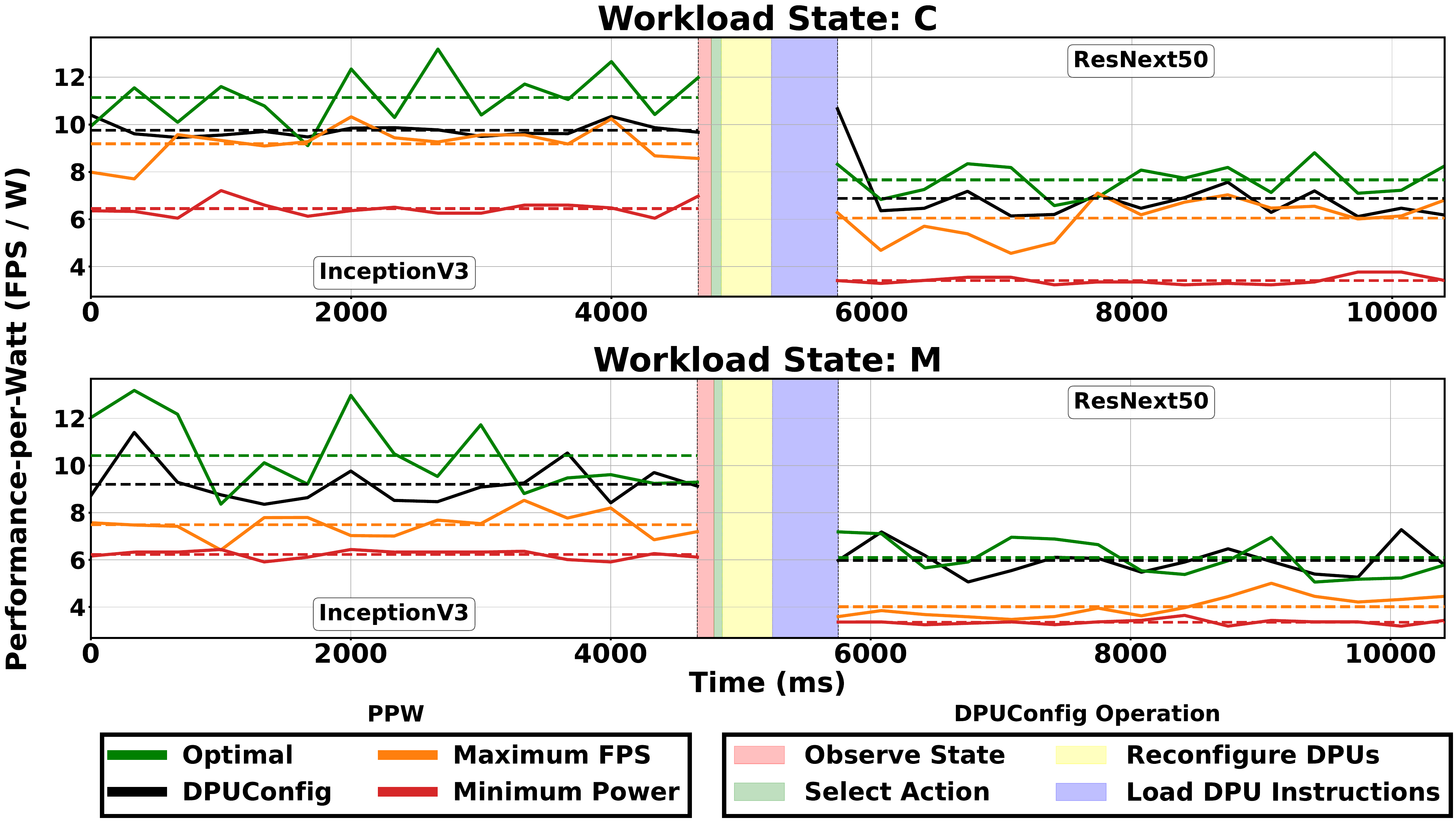}
    \caption{Timeline of DPUConfig operation during inference of InceptionV3 and ResNext50, where a reconfiguration takes place.} 
    \label{fig:timeline}
\end{figure}

Fig.~\ref{fig:timeline} shows a timeline in which a new model arrives and a DPU configuration is selected. The solid lines represent the observed PPW during InceptionV3 and ResNext50 inference, while the dashed lines indicate the average PPW. We compare with the same baselines as in Fig.~\ref{fig:workload_fig}. The shaded regions illustrate the overheads measured on ZCU102: telemetry collection for state observation (88\,ms), RL inference on the Arm CPU for action selection (20\,ms), DPU reconfiguration (384\,ms), and instruction loading (507\,ms). In this snapshot, the DPU changes, so all phases are included. If the same DPU is reused, reconfiguration and loading are not needed. For long-running deep neural network inferences, the overhead of about 1047\,ms is negligible compared to the inference runtime, while enabling near-optimal efficiency.

\section{Prior work}
\label{PriorWork}

\textbf{CNN accelerators.} 
Many works propose efficient CNN accelerators for FPGAs~\cite{guo2019dl, umuroglu2017finn, fahim2021hls4ml}. Specifically for DPU design, Du et al.~\cite{DuZZSJ23} propose a framework for accelerating DNN inference on FPGAs using heterogeneous multi-DPU engines, extending Vitis AI support for homogeneous DPU deployments. Their system uses both task-level and pipelined parallelism to map CNN layers to distinct DPUs based on resource profiles. The method improves inference throughput by up to 19\% on the ZCU104 platform, improving resource utilization and scheduling efficiency for CNN workloads. Unlike \framework, this work does not employ RL methods for optimization. Also, emphasis is purely on improving performance, rather than achieving energy efficiency.

EXPRESS~\cite{express_tecs2024} is an ML-based framework that accurately predicts execution time for CNNs running concurrently on Xilinx DPU accelerators. EXPRESS and its improved version, EXPRESS-2.0, incorporate hardware, model, and interconnect features to reduce prediction error to as low as 0.7\%, enabling efficient design space exploration targeting multi-CNN deployments on FPGAs. This work focuses on predicting the latency of CNNs running on DPUs, rather than suggesting methods to configure DPUs to optimise their energy efficiency.

\textbf{RL for system optimization.} Recent work applies RL to system optimization on heterogeneous platforms. ConfuciuX~\cite{kao2020confuciux} frames per-layer allocation of PEs and buffers in DNN accelerators under area/power limits as an RL problem, then uses a genetic algorithm for fine-tuning, converging 4.7×–24× faster than prior methods. BAND~\cite{li2022band} dynamically schedules multi-DNN workloads via fine-grained subgraph placement across CPUs, GPUs, and NPUs to reduce makespan latency. HERTI~\cite{liu2021herti} similarly employs an RL scheduler with compute/communication cost estimators for model-slice scheduling. More recent work extends DL scheduling to include FPGAs~\cite{wu2025reinforcement}. This prior work uses RL for task scheduling, but not for DPU configuration as \framework does.
\section{Conclusions}
\label{conclusion}

To support energy-efficient ML inference in an FPGA-based MPSoC platform, we introduce \framework, a custom RL-based agent that continuously learns and selects near-optimal DPU configurations by taking into account the features of the ML model and dynamically adapting to the stochastic runtime variance. The evaluations conducted on a ZCU102 MPSoC show that, on average, \framework selects the best DPU configuration 95\% of the optimal PPW, proving the effectiveness of our framework. 

\bibliographystyle{IEEEtran}
\bibliography{99_references}

@misc{otel,
  key =          {otel},
  title =        {{Open Telemetry}},
  howpublished = {\url{https://opentelemetry.io}}
}

@misc{xilinx_zcu102,
  author       = {{Xilinx Inc.}},
  title        = {{Zynq UltraScale+ MPSoC ZCU102 Evaluation Kit}},
  howpublished = {\url{https://www.xilinx.com/products/boards-and-kits/zcu102.html}},
  year         = {2021}
}

@manual{amd2023vitisai,
  title        = {{Vitis AI User Guide (UG1414, Version 3.5)}},
  author       = {{AMD/Xilinx Inc.}},
  year         = {2023},
  month        = sep
}

@manual{xilinx_dpuczdx8g,
  title        = {{DPUCZDX8G for Zynq UltraScale+ MPSoCs: Product Guide}},
  author       = {{Xilinx Inc.}},
  organization = {Xilinx},
  year         = {2021},
  note         = {PG338 (v3.2), Accessed: 2025-06-29},
}

@article{mittal2020survey,
  author    = {Sudeep Mittal},
  title     = {A survey of {{FPGA-based}} Accelerators for Convolutional Neural Networks},
  journal   = {Neural Computing and Applications},
  volume    = {32},
  number    = {4},
  pages     = {1109--1139},
  year      = {2020},
  publisher = {Springer},
  doi       = {10.1007/s00521-018-3970-5}
}

@misc{yan2024surveyfpgabasedacceleratorml,
      title={{A survey on {{FPGA-based}} Accelerator for ML Models}}, 
      author={Feng Yan and Andreas Koch and Oliver Sinnen},
      year={2024},
      eprint={2412.15666},
      archivePrefix={arXiv},
      primaryClass={cs.AR},
      url={https://arxiv.org/abs/2412.15666}, 
}

@article{hussein2022thermal,
  title     = {{Implementation of a {{DPU-Based}} Intelligent Thermal Imaging Hardware Accelerator on FPGA}},
  author    = {Hussein, Mohamed S. and Soliman, Mohamed A. and Elhoseny, Mohamed and Khalil, Amr and Wahba, Khaled},
  journal   = {Electronics},
  volume    = {11},
  number    = {1},
  pages     = {105},
  year      = {2022},
  publisher = {MDPI}
}

@inproceedings{wu2019autonomous,
  author    = {T. Wu and W. Liu and Y. Jin},
  title     = {{An End-to-End Solution to Autonomous Driving Based on Xilinx FPGA}},
  booktitle = {2019 International Conference on Field-Programmable Technology (ICFPT)},
  year      = {2019},
  pages     = {427--430},
  publisher = {IEEE},
  doi       = {10.1109/ICFPT47387.2019.00075}
}

@article{Li2023EdgeRealTimeDPU,
  author       = {Li, Chao and Xu, Rui and Lv, Yong and Zhao, Yonghui and Jing, Weipeng},
  title        = {{Edge Real‑Time Object Detection and DPU‑Based Hardware Implementation for Optical Remote Sensing Images}},
  journal      = {Remote Sensing},
  year         = {2023},
  volume       = {15},
  number       = {16},
  pages        = {3975},
  doi          = {10.3390/rs15163975},
  url          = {https://doi.org/10.3390/rs15163975}
}

@article{Amin2024FPGARealTime,
  author       = {Amin, Rashed Al and Hasan, Mehrab and Wiese, Veit and Obermaisser, Roman},
  title        = {{FPGA‑based Real‑Time Object Detection and Classification System using YOLO for Edge Computing}},
  journal      = {IEEE Access},
  year         = {2024},
  volume       = {PP(99)},
  doi          = {10.1109/ACCESS.2024.3404623},
}

@article{Perryman2025DependableDPU,
  author       = {Perryman, Noah and Sabogal, Sebastian and Wilson, Christopher and George, Alan D.},
  title        = {{Dependable DPU Architectures on AMD‑Xilinx Versal Adaptive SoCs for Space Applications}},
  journal      = {IEEE Transactions on Aerospace and Electronic Systems},
  year         = {2025} 
}

@inproceedings{He2016ResNet,
  author       = {Kaiming He and Xiangyu Zhang and Shaoqing Ren and Jian Sun},
  title        = {{Deep Residual Learning for Image Recognition}},
  booktitle    = {Proceedings of the IEEE Conference on Computer Vision and Pattern Recognition (CVPR)},
  pages        = {770--778},
  year         = {2016},
  month        = jun,
  publisher    = {IEEE},
  doi          = {10.1109/CVPR.2016.90},
  location     = {Las Vegas, NV, USA}
}

@inproceedings{Sandler2018MobileNetV2,
  author    = {Mark Sandler and Andrew G. Howard and Menglong Zhu and Andrey Zhmoginov and Liang‑Chieh Chen},
  title     = {{MobileNetV2: Inverted Residuals and Linear Bottlenecks}},
  booktitle = {Proceedings of the IEEE Conference on Computer Vision and Pattern Recognition (CVPR)},
  pages     = {4510--4520},
  month     = jun,
  year      = {2018},
  publisher = {IEEE},
  doi       = {10.1109/CVPR.2018.00474}
}

@inproceedings{Deng2009ImageNet,
  author    = {Jia Deng and Wei Dong and Richard Socher and Li‑Jia Li and Kai Li and Li Fei‑Fei},
  title     = {{ImageNet: A Large‑Scale Hierarchical Image Database}},
  booktitle = {Proceedings of the 2009 IEEE Conference on Computer Vision and Pattern Recognition (CVPR)},
  pages     = {248--255},
  year      = {2009},
  month     = jun,
  publisher = {IEEE},
  doi       = {10.1109/CVPR.2009.5206848},
}

@book{sutton2018reinforcement,
  title={{Reinforcement Learning: An Introduction}},
  author={Sutton, Richard S. and Barto, Andrew G.},
  year={2018},
  edition={2nd},
  publisher={A Bradford Book},
  address={Cambridge, MA, USA}
}

@inproceedings{kao2020confuciux,
  author       = {Kao, Sheng‐Chun and Jeong, Geonhwa and Krishna, Tushar},
  title        = {{ConfuciuX: Autonomous Hardware Resource Assignment for DNN Accelerators using Reinforcement Learning}},
  booktitle    = {Proceedings of the 53rd Annual IEEE/ACM International Symposium on Microarchitecture (MICRO)},
  year         = {2020},
  pages        = {622--636},
  publisher    = {IEEE}
}

@inproceedings{liu2021herti,
  title={{HERTI: A Reinforcement Learning-Augmented System for Efficient Real-Time Inference on Heterogeneous Embedded Systems}},
  author={Liu, Zhe and Wu, Xiaojun and Huang, Kai},
  booktitle={{Proceedings of the 2021 International Conference on Parallel Architectures and Compilation Techniques (PACT)}},
  pages={195--207},
  year={2021},
  organization={IEEE},
  doi={10.1109/PACT52795.2021.00024}
}

@inproceedings{li2022band,
  title={{BAND: Coordinated Multi-DNN Inference on Heterogeneous Mobile Processors}},
  author={Li, Xiaohua and Zhu, Yuhao and Zhang, Yiying},
  booktitle={Proceedings of the 20th Annual International Conference on Mobile Systems, Applications, and Services (MobiSys)},
  pages={175--188},
  year={2022},
  organization={ACM},
  doi={10.1145/3498361.3538934}
}

@article{wu2025reinforcement,
  title={{Reinforcement Learning-Based Task Scheduling for Heterogeneous Computing in End-Edge-Cloud Environment}},
  author={Wu, Yuxiang and Wu, Guowei and Wang, Yingwei and Liao, Xiaofei},
  journal={Cluster Computing},
  year={2025},
  publisher={Springer}
}

@inproceedings{DuZZSJ23,
  author    = {Zelin Du and
               Wei Zhang and
               Zimeng Zhou and
               Zili Shao and
               Lei Ju},
  title     = {{Accelerating DNN Inference with Heterogeneous Multi-DPU Engines}},
  booktitle = {Proceedings of the 60th Design Automation Conference (DAC)},
  pages     = {1--6},
  year      = {2023}
}

@article{express_tecs2024,
  title     = {{EXPRESS: A Framework for Execution Time Prediction of Concurrent CNNs on Xilinx DPU Accelerator}},
  author    = {S. Goel and R. Kedia and R. Sen and M. Balakrishnan},
  journal   = {ACM Transactions on Embedded Computing Systems (TECS)},
  year      = {2024},
  doi       = {10.1145/3697835}
}

@misc{densenet,
      title={Densely Connected Convolutional Networks}, 
      author={Gao Huang and Zhuang Liu and Laurens van der Maaten and Kilian Q. Weinberger},
      year={2018},
      eprint={1608.06993},
      archivePrefix={arXiv},
      primaryClass={cs.CV},
      url={https://arxiv.org/abs/1608.06993}, 
}

@misc{regnet,
      title={Designing Network Design Spaces}, 
      author={Ilija Radosavovic and Raj Prateek Kosaraju and Ross Girshick and Kaiming He and Piotr Dollár},
      year={2020},
      eprint={2003.13678},
      archivePrefix={arXiv},
      primaryClass={cs.CV},
      url={https://arxiv.org/abs/2003.13678}, 
}

@misc{resnext,
      title={Aggregated Residual Transformations for Deep Neural Networks}, 
      author={Saining Xie and Ross Girshick and Piotr Dollár and Zhuowen Tu and Kaiming He},
      year={2017},
      eprint={1611.05431},
      archivePrefix={arXiv},
      primaryClass={cs.CV},
      url={https://arxiv.org/abs/1611.05431}, 
}

@inproceedings{umuroglu2017finn,
  author       = {{Umuroglu, Yaman and others}},
  title        = {{FINN: A Framework for Fast, Scalable Binarized Neural Network Inference}},
  booktitle    = {Proceedings of the 2017 ACM/SIGDA International Symposium on Field-Programmable Gate Arrays (FPGA)},
  year         = {2017},
  pages        = {65--74},
  publisher    = {ACM}
}

@article{fahim2021hls4ml,
  author       = {Farah Fahim and
                  others},
  title        = {{hls4ml: An Open-Source Codesign Workflow to Empower Scientific Low-Power Machine Learning Devices}},
  journal      = {CoRR},
  volume       = {abs/2103.05579},
  year         = {2021},
  url          = {https://arxiv.org/abs/2103.05579},
  eprinttype    = {arXiv},
  eprint       = {2103.05579}
}

@article{guo2019dl,
  title={{[DL] A survey of FPGA-based neural network inference accelerators}},
  author={Guo, Kaiyuan and Zeng, Shulin and Yu, Jincheng and Wang, Yu and Yang, Huazhong},
  journal={ACM Transactions on Reconfigurable Technology and Systems (TRETS)},
  volume={12},
  number={1},
  pages={1--26},
  year={2019},
  publisher={ACM New York, NY, USA}
}

@inproceedings{liContextualBandit,
  title = {A {{Contextual-Bandit Approach}} to {{Personalized News Article Recommendation}}},
  booktitle = {Proceedings of the 19th International Conference on {{World}} Wide Web},
  author = {Li, Lihong and Chu, Wei and Langford, John and Schapire, Robert E.},
  year = {2010},
  month = apr,
  eprint = {1003.0146},
  primaryclass = {cs},
  pages = {661--670},
  doi = {10.1145/1772690.1772758},
  urldate = {2025-09-08},
  archiveprefix = {arXiv}
}

@inproceedings{maoVarianceReductionReinforcement2018,
  author       = {Hongzi Mao and
                  Shaileshh Bojja Venkatakrishnan and
                  Malte Schwarzkopf and
                  Mohammad Alizadeh},
  title        = {{Variance Reduction for Reinforcement Learning in Input-Driven Environments}},
  booktitle    = {7th International Conference on Learning Representations, {ICLR} 2019,
                  New Orleans, LA, USA, May 6-9, 2019},
  year         = {2019}
}

@inproceedings{ngPolicyInvarianceReward1999,
  title = {Policy {{Invariance Under Reward Transformations}}: {{Theory}} and {{Application}} to {{Reward Shaping}}},
  shorttitle = {Policy {{Invariance Under Reward Transformations}}},
  booktitle = {{16th International Conference on Machine Learning (ICML)}},
  author = {Ng, Andrew Y. and Harada, Daishi and Russell, Stuart J.},
  year = {1999},
  month = jun,
  series = {{{ICML}} '99},
  pages = {278--287},
  address = {San Francisco, CA, USA},
  urldate = {2025-09-08},
}

@misc{tuckerMirageActionDependentBaselines2018,
  title = {The {{Mirage}} of {{Action-Dependent Baselines}} in {{Reinforcement Learning}}},
  author = {Tucker, George and Bhupatiraju, Surya and Gu, Shixiang and Turner, Richard E. and Ghahramani, Zoubin and Levine, Sergey},
  year = {2018},
  month = nov,
  number = {arXiv:1802.10031},
  eprint = {1802.10031},
  primaryclass = {cs},
  doi = {10.48550/arXiv.1802.10031},
  urldate = {2025-09-05},
  archiveprefix = {arXiv},
  langid = {english}
}

@online{fuRewardShapingMitigate2025,
  title = {Reward {{Shaping}} to {{Mitigate Reward Hacking}} in {{RLHF}}},
  author = {Fu, Jiayi and Zhao, Xuandong and Yao, Chengyuan and Wang, Heng and Han, Qi and Xiao, Yanghua},
  date = {2025-06-17},
  eprint = {2502.18770},
  eprinttype = {arXiv},
  eprintclass = {cs},
  doi = {10.48550/arXiv.2502.18770},
  url = {http://arxiv.org/abs/2502.18770},
  urldate = {2025-09-05},
  langid = {english},
  pubstate = {prepublished},
}

@inproceedings{hesselRainbowCombiningImprovements2018,
  title = {Rainbow: Combining Improvements in Deep Reinforcement Learning},
  year = {2018},
  shorttitle = {Rainbow},
  booktitle = {32nd Conference on Artificial Intelligence},
  author = {Hessel, Matteo and others},
  date = {2018-02-02},
  pages = {3215--3222},
  publisher = {AAAI Press},
  location = {New Orleans, Louisiana, USA},
  isbn = {978-1-57735-800-8}
}

@article{mnihHumanlevelControlDeep2015,
  title = {{Human-Level Control through Deep Reinforcement Learning}},
  author = {Mnih, Volodymyr and others},
  year = {2015},
  date = {2015-02-26},
  journaltitle = {Nature},
  shortjournal = {Nature},
  volume = {518},
  number = {7540},
  pages = {529--533},
  urldate = {2025-09-08},
  langid = {english}
}

@online{schulmanPPO2017,
  title = {Proximal {{Policy Optimization Algorithms}}},
  author = {Schulman, John and Wolski, Filip and Dhariwal, Prafulla and Radford, Alec and Klimov, Oleg},
  date = {2017-08-28},
  eprint = {1707.06347},
  eprinttype = {arXiv},
  eprintclass = {cs},
  doi = {10.48550/arXiv.1707.06347},
  url = {http://arxiv.org/abs/1707.06347},
  urldate = {2025-09-08},
  pubstate = {prepublished},
}

@article{EagleEye2020,
  author       = {Bailin Li and
                  Bowen Wu and
                  Jiang Su and
                  Guangrun Wang and
                  Liang Lin},
  title        = {{EagleEye: Fast Sub-net Evaluation for Efficient Neural Network Pruning}},
  journal      = {CoRR},
  volume       = {abs/2007.02491},
  year         = {2020},
  url          = {https://arxiv.org/abs/2007.02491},
  eprinttype    = {arXiv},
  eprint       = {2007.02491},
  bibsource    = {dblp computer science bibliography, https://dblp.org}
}

@misc{lin2015microsoftcococommonobjects,
      title={Microsoft COCO: Common Objects in Context}, 
      author={Tsung-Yi Lin and Michael Maire and Serge Belongie and Lubomir Bourdev and Ross Girshick and James Hays and Pietro Perona and Deva Ramanan and C. Lawrence Zitnick and Piotr Dollár},
      year={2015},
      eprint={1405.0312},
      archivePrefix={arXiv},
      primaryClass={cs.CV},
      url={https://arxiv.org/abs/1405.0312}, 
}

@inproceedings{pytorch,
author = {Ansel, Jason and Yang, et. al.},
title = {PyTorch 2: Faster Machine Learning Through Dynamic Python Bytecode Transformation and Graph Compilation},
year = {2024},
isbn = {9798400703850},
publisher = {Association for Computing Machinery},
address = {New York, NY, USA},
url = {https://doi.org/10.1145/3620665.3640366},
doi = {10.1145/3620665.3640366},
pages = {929–947},
numpages = {19},
location = {La Jolla, CA, USA},
series = {ASPLOS '24}
}

@misc{onnx,
    author = {Bai, Junjie and Lu, Fang and Zhang, Ke and others},
    title = {{ONNX}: Open Neural Network Exchange},
    year = {2019},
    publisher = {GitHub},
    journal = {GitHub repository},
    howpublished = {\url{https://github.com/onnx/onnx}},
    commit = {94d238d96e3fb3a7ba34f03c284b9ad3516163be}
}

@misc{king2025stressng,
  title        = {stress-ng},
  howpublished = {\url{https://github.com/ColinIanKing/stress-ng}},
  year         = {2025},
  note         = {Version 0.19.04}
}

@misc{nodexporter,
  title = {Prometheus Node Exporter},
  year = {2025},
  month = sep,
  urldate = {2025-09-14},
  howpublished = {\url{https://github.com/prometheus/node\_exporter}}
}

@article{brockman2016openai,
  title={OpenAI Gym},
  author={Brockman, Greg and Cheung, Vicki and Pettersson, Ludwig and Schneider, Jonas and Schulman, John and Tang, Jie and Zaremba, Wojciech},
  journal={arXiv preprint arXiv:1606.01540},
  year={2016}
}

@inproceedings{raylib,
author = {Moritz, Philipp and Nishihara, Robert and Wang, Stephanie and Tumanov, Alexey and Liaw, Richard and Liang, Eric and Elibol, Melih and Yang, Zongheng and Paul, William and Jordan, Michael I. and Stoica, Ion},
title = {Ray: a distributed framework for emerging AI applications},
year = {2018},
isbn = {9781931971478},
publisher = {USENIX Association},
address = {USA},
booktitle = {Proceedings of the 13th USENIX Conference on Operating Systems Design and Implementation},
pages = {561–577},
numpages = {17},
location = {Carlsbad, CA, USA},
series = {OSDI'18}
}

\end{document}